# A Real-Time BCI for Stroke Hand Rehabilitation Using Latent EEG Features from Healthy Subjects

1st Fady.M..Omar
Department Of Mechatronics Engineering
Military Technical College
Cairo,Egypt
Eng.fady99@gmail.com

2nd Amr.M.Omar
Department Of Neuroscience,School of Medicine
Queen's university
Ontario,Canada
Dr.Amr.omar99@gmail.com

3nd Khaled Eyada
Department Of Electronics and Comunication
Seuz Canal university
Ismailia,Egypt
Khaledeyada@gmail.com

4th Mohamed Rabie
Department Of Biomedical Engineering
Military Technical College
Cairo, Egypt
mohamed.rabie@mtc.edu.eg

5th Ahmed.m.azab
Department Of Biomedical Engineering
Military Technical College
Cairo, Egypt
Ahmed.m.azab@ieee.org

6th Mohamed A. Kamel
Department Of Mechatronics Engineering,
Military Techncial College,
Cairo, Egypt
mohamed.atef.kamel@mtc.edu.eg

*Abstract*— This study presents a real-time, portable brain–computer interface (BCI) system designed to support hand rehabilitation for stroke patients. The system combines a low-cost 3D-printed robotic exoskeleton with an embedded controller that converts brain signals into physical hand movements. EEG signals are recorded using a 14-channel Emotiv EPOC+ headset and processed through a supervised convolutional autoencoder (CAE) to extract meaningful latent features from single-trial data. The model is trained on publicly available EEG data from healthy individuals (WAY-EEG-GAL dataset), with electrode mapping adapted to match the Emotiv headset layout. Among several tested classifiers, AdaBoost achieved the highest accuracy (89.3%) and F1-score (0.89) in offline evaluations. The system was also tested in real time on five healthy subjects, achieving classification accuracies between 60% and 86%. The complete pipeline—EEG acquisition, signal processing, classification, and robotic control—is deployed on an NVIDIA Jetson Nano platform with a real-time graphical interface. These results demonstrate the system's potential as a low-cost, standalone solution for home-based neurorehabilitation.

*Keywords — Brain Computer Interface (BCI), EEG Classification, Neurorehabilitation, Convolutional Autoencoder, Robotic Exoskeleton*

## I. INTRODUCTION

Brain–Computer Interface (BCI) systems allow the brain to communicate directly with external devices, which makes them especially promising for neurorehabilitation [1]. Among the different types of BCIs, EEG-based systems stand out because they're affordable, portable, and non-invasive— qualities that make them ideal for decoding motor intentions in people with movement difficulties [1].One of the main challenges in building these systems is reliably detecting different hand movement states, especially distinguishing between rest and activity. Recent research has shown that deep learning methods can significantly improve how EEG signals are processed by capturing both spatial and temporal features [2]. In particular, brain patterns like movement-related cortical potentials (MRCPs) and event-related desynchronization/synchronization (ERD/ERS) have proven useful for identifying motor activity more accurately [3].To work effectively in real-life rehabilitation settings, BCI systems need to respond in real time with very low delay. Some studies have shown that it's possible to run EEG-based BCIs on small embedded devices with latencies under 50 milliseconds, allowing for smooth and responsive control [4]. This is especially valuable in stroke rehabilitation, where repeating intended or imagined movements can help the brain form new connections. Adding robotic exoskeletons to the loop makes this feedback more engaging and supports motor learning [5]. Our earlier work explored the use of deep convolutional neural networks (CNNs) to control a prosthetic arm using EEG signals, demonstrating the potential of brain-driven assistive devices for real-time applications [6]. Building on that foundation, this study shifts the focus from prosthetics to rehabilitation, proposing a complete system that supports hand recovery through brain-controlled robotic feedback. To improve signal interpretation, many studies have used a mix of neural networks and signal processing techniques—such as CNN-LSTM models, Common Spatial Patterns (CSP), and wavelet transforms—to improve accuracy in classifying motor states [7]. More advanced approaches, like attention-based models and tripolar EEG setups, have shown potential in improving signal quality and reducing variability between users [8]. Finally, integrating these models into practical, affordable systems—using consumer-grade EEG headsets and optimized classification pipelines—has made real-time BCI control increasingly feasible [9]. These advances suggest that it is now possible to build portable, low-cost BCI systems that support hand movement rehabilitation. However, many existing systems still lack real-time capability or require complex equipment. In this work, we present and evaluate a complete, real-time BCI system aimed at hand rehabilitation for stroke patients. Our system is trained and validated using EEG data from healthy individuals.

## II. RELATED WORK

In recent years, EEG-based brain–computer interface (BCI) systems have become more accurate and practical for neurorehabilitation, particularly in decoding motor intentions. A key trend in this field is the ability to interpret more detailed aspects of movement, such as direction and speed. For example, researchers have used wavelet features and MRCP signals to classify bi-directional and velocity-specific motor imagery, achieving over 68% accuracy across users [10].Deep learning models that integrate physiological features—such as movement-related cortical potentials (MRCPs) and event-related desynchronization/synchronization (ERD/ERS)—have shown impressive results in decoding both single-hand and bimanual movements. These models often use convolutional layers and attention mechanisms to extract strong and meaningful EEG features suitable for rehabilitation-focused applications [3].Other studies have demonstrated that combining kinesthetic motor imagery (KMI) with simple visual cues can enhance classification performance, especially when used with robotic orthoses in stroke rehabilitation. Adding active movement support and tuning the spectral features further improved accuracy by about 3% compared to standard methods [5].One promising approach involves using action observation (AO) to activate the brain's mirror neuron system. In one study, providing users with anticipated visual feedback during AO significantly altered mu and beta rhythms, helping increase engagement and stimulate brain plasticity during recovery tasks [11]. Also, the way feedback is delivered matters—continuous robotic feedback has been found to produce better BCI accuracy and stronger cortical activation than discrete feedback, likely due to improved sensorimotor integration [12].To expand system capabilities, some researchers have developed multimodal platforms that combine EEG with virtual reality (VR) and wearable multi-degree-of-freedom (multi-DOF) robotic devices. These systems offer immersive, real-time feedback and use advanced classifiers like LightGBM to achieve high accuracy by fusing spectral and temporal EEG features [13].In addition, the availability of high-quality datasets is helping accelerate progress. For instance, the MOVING dataset provides synchronized EEG and kinematic recordings for various hand gestures, supporting the training and benchmarking of deep learning models such as EEGNetV4 for assistive movement decoding [14]. In response to these advances, our work presents a portable, real-time BCI system designed specifically for hand rehabilitation. The system combines motor imagery (MI) and motor execution (ME) decoding using a supervised convolutional autoencoder (CAE) that performs both feature learning and signal reconstruction. The compact latent features are then classified using an AdaBoost classifier. The model is trained offline using EEG data from healthy individuals in the WAY-EEG-GAL dataset. To enable real-time testing, we mapped the training electrodes to match the 14-channel Emotive EPOC+ headset. The system runs entirely on an NVIDIA Jetson Nano, and decoded motor intentions are used to control a 3D-printed robotic exoskeleton hand that provides both proprioceptive and visual feedback.as shown as in figure 1.This work makes four key contributions:

First, A supervised CAE model that learns compact and discriminative EEG features, outperforming conventional baselines in offline tests. Second, A practical strategy for transferring models from lab-grade datasets to consumer-grade EEG devices using electrode mapping. Third, A fully integrated, low-latency embedded pipeline—including EEG acquisition, preprocessing, feature extraction, classification, and robotic actuation—on a Jetson Nano. Forth, Real-time validation on five healthy participants, demonstrating consistent classification accuracy (60–86%) and reliable control of the robotic exoskeleton.

## III. MATERIALS AND METHODS

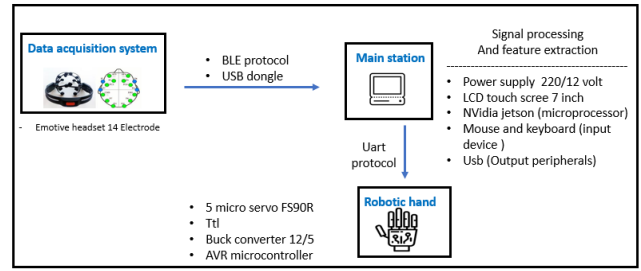

Figure.1. *Main block diagram for our Rehabilitation System*

### A. Dataset and Participants

To develop and evaluate the proposed BCI system, we used EEG data from healthy individuals in both offline and real-time settings. For offline training and validation, we relied on the WAY-EEG-GAL public dataset [15], which contains EEG recordings of healthy subjects performing grasp-and-lift tasks. Specifically, we selected 1,495 trials labeled as FirstDigitTouch to represent the active hand movement class. Since the dataset doesn't include a clear "rest" label, we extracted rest-state samples from the inter-trial intervals, using timing information and event markers to isolate moments when the hand was in a neutral position. The original EEG data were recorded using a 32-channel cap based on the international 10–20 system and sampled at 256 Hz. To make the model compatible with our real-time system, which uses the 14-channel Emotiv EPOC+ headset, we selected 12 overlapping electrodes during preprocessing for training and validation (see Figure 2). For online testing, we recruited five healthy adult volunteers (aged 24 to 32; 3 males and 2 females). These participants wore the Emotiv EPOC+ headset and performed motor imagery tasks involving hand movement. The real-time EEG data collected from them were used to evaluate system performance under realistic conditions after deploying the trained model.

### B. Signal Acquisition and Preprocessing

EEG signals were acquired and processed using a consistent pipeline applied in both the offline training and real-time testing stages. Twelve scalp electrodes (F7, F3, FC5, T7, P7, O1, O2, P8, T8, FC6, F4, F8) were chosen as shown as in figure 2. These channels were selected as they are present in both the WAY-EEG-GAL data set and the Emotiv EPOC+ headset configuration in order to facilitate model transfer between data sets. All EEG recordings were down sampled, or unsampled, at the 250 Hz sampling rate to obtain equal sample rate in time for all the sessions. Each trial was segmented into 1-second windows, resulting in 250 time points per trial. Care was taken to preserve the spatial alignment of electrode positions to maintain physiological relevance during analysis. To clean the signals, a bandpass filter between 8 and 40 Hz was applied. This frequency band captures activity related to motor intention while removing slow drifts and high-frequency noise. Line noise at 50/60 Hz was already filtered out in the original dataset. In addition to filtering, we used Independent Component Analysis (ICA) to remove non-brain artifacts. Components identified as originating from muscle movements, eye blinks, or cardiac signals were excluded based on their spatial distribution and temporal profiles. Only the components reflecting neural activity were retained for further processing. This preprocessing workflow ensured that the EEG input to our feature extraction and classification pipeline was clean, consistent, and representative of true brain activity related to hand movement tasks.

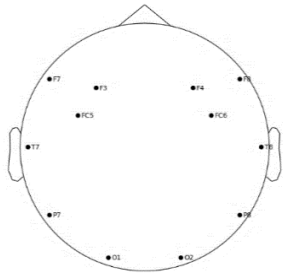

Figure.1. *The topographical layout of the 12 EEG electrodes selected for analysis corresponds to overlapping positions between the WAY-EEG-GAL dataset and the Emotiv EPOC+ headset.*

## C. Feature Extraction

We propose a supervised convolutional autoencoder combined with a lightweight boosted classifier, referred to as SupCAE-Boost (SCB), designed for efficient and real-time decoding of EEG signals. Each EEG window consists of 12 channels with 250 time samples (12×250), representing a short segment of neural activity as shown as in figure 3. The encoder transforms each of these windows into a compact "summary" vector that captures the most relevant temporal–spatial features of the signal. This representation is then passed to a boosted classifier, which performs the final discrimination between movement and rest states in real time. The encoder itself is composed of three sequential convolutional layers (32, 64, and 128 filters), each followed by batch normalization and a leaky-ReLU activation to stabilize and accelerate training. Pooling layers are inserted to gradually reduce the temporal resolution, allowing the network to focus attention on higher-level patterns while suppressing noise and redundancy. Ultimately, each EEG window is compressed into a 64-dimensional latent vector. This latent representation is intentionally compact to ensure rapid execution on embedded hardware, yet sufficiently expressive to retain the discriminative features needed for reliable classification. Training the model is formulated as a multi-objective task. The first objective is to reconstruct the original EEG signal from the latent space, ensuring that the encoder preserves the essential signal characteristics. The second objective is to encourage separation between movement and rest states so that the latent features become directly useful for classification. To achieve this, a small auxiliary classifier head is attached to the latent vector during training. At inference, this auxiliary head is removed, leaving only the encoder paired with the boosted classifier. Regularization techniques, including dropout, weight decay, and early stopping, are employed to reduce overfitting, while lightweight augmentations such as Gaussian noise injection and random channel dropout further improve robustness against variability in EEG recordings. The boosted classifier is chosen as the final decision stage because it provides an excellent balance between accuracy, speed, and resource efficiency. In particular, ensemble methods such as AdaBoost are well suited to operate on compact latent features, offering robustness with limited training data, very low memory requirements, and high stability across sessions. For deployment, the encoder is exported in reduced numerical precision (FP16/INT8) and executed on a Jetson Nano, where it processes each EEG window in only a few milliseconds. This ensures that the system remains responsive and suitable for real-time neurorehabilitation applications.

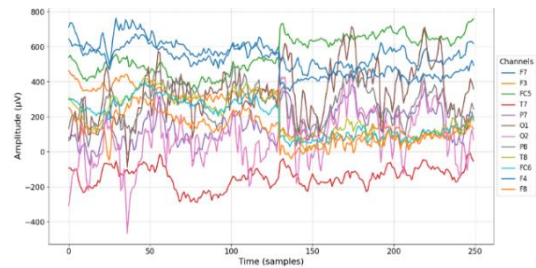

Figure.3. *Raw EEG signals from 12 electrodes of Subject 1 in the dataset comprising 250 samples (1 second) before any artifact removal.*

## D. Classification

Latent features extracted from the convolutional autoencoder were used to train a range of classifiers, including traditional machine learning models and deep neural architectures. Classical methods such as SVM, LDA, decision trees, and ensemble-based classifiers were evaluated alongside shallow feedforward networks, recurrent models, and convolutional architectures. All models were trained on the same low-dimensional feature space and validated using a fixed data split. The classifier demonstrating the best generalization performance was selected for integration into the real-time system. To ensure fair and reproducible comparisons, all baselines are trained and evaluated on the same CAE latent embeddings and identical data splits, with common preprocessing and label definitions. Hyperparameters are selected via cross-validation on training folds only; subject-independent performance is assessed with LOSO without leakage. We report window- and trial-level Accuracy, F1, and Macro-F1 with 95% confidence intervals, and measure inference latency, memory, and energy per decision on a Jetson Nano.

## E. Robotic Exoskeleton Hand Design

The hand exoskeleton was designed for low mass, modularity, and safe assisted motion suitable for rehabilitation use. The structural frame is 3D-printed from PETG for stiffness and durability, while all skin-contact interfaces are printed in TPU to provide compliant support and distribute pressure. A dorsal mounting arrangement with soft straps stabilizes the device without obstructing the palmar side. Each finger module is independently actuated by a micro-servo (FS90MG, 180°, metal gears; stall torque 2.2 kg • cm ≈ 21.6 N • cm), driving a lightweight linkage to assist flexion/extension. The total device mass is 450 g, including cabling and fasteners. The nominal range of motion per finger is 120°, suitable for slow, repeatable therapeutic movements.

*F. Embedded Real-Time Control System*

A dedicated embedded system was developed to enable real-time neurorehabilitation control. The trained BCI model was deployed on an NVIDIA Jetson Nano module (4 GB RAM), which served as the core processor for EEG decoding and classification. A custom-designed graphical user interface (GUI) displayed a virtual hand with open and closed states, providing both visual feedback and user interaction. The decoded command was transmitted to a secondary control board containing a microcontroller responsible for driving the servo motors embedded in the exoskeleton. The system supported both active user intention (via BCI) and passive assistance modes, ensuring adaptability to varying levels of motor ability. All components were fully integrated to maintain low latency, portability, and operational autonomy.

## IV. RESULTS AND EVALUATION

*A. EEG Signal Before and After Preprocessing*

ICA was employed to remove non-EEG components, including those related to heartbeat (Components 0 and 3), muscle activity (Component 1), and other non-neural sources (Component 11), as shown in Fig. 4. Neural components were retained for further analysis. To enhance signal fidelity, a bandpass filter between 8–40 Hz was applied, targeting the frequency range most relevant to motor-related activity, particularly hand and finger movements, which predominantly manifest within this spectrum. This range includes the mu (8–13 Hz) and beta (13–30 Hz) rhythms, as well as part of the low-gamma band, all of which are closely linked to motor activity. Earlier studies have shown that using the 8–40 Hz range provides more stable and accurate results than restricting the filter to only the mu and beta bands [16]. As illustrated in Fig. 5, the resulting signal from channel F4 demonstrates a marked reduction in artifacts, highlighting the effectiveness of the preprocessing pipeline in isolating physiologically meaningful EEG patterns.

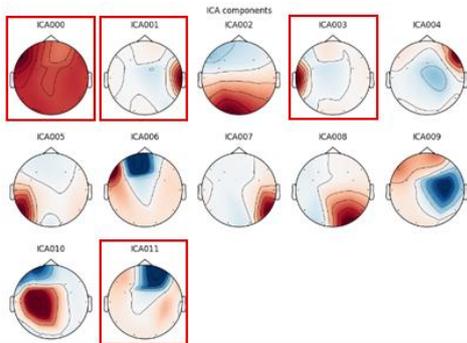

Figure.4. *ICA decomposition of EEG signals identifies non-EEG components related to heartbeat, muscle activity, and other sources.*

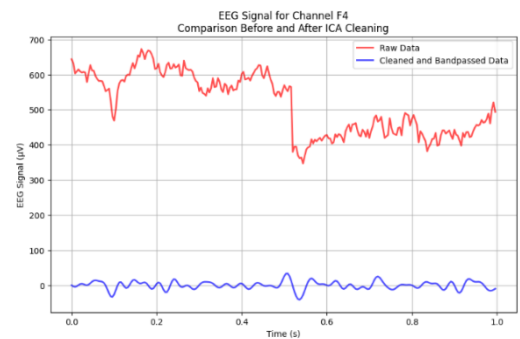

Figure.5. *EEG signal from channel F4 before and after ICA-based artifact removal and bandpass filtering (8 and 40 Hz).*

*B. Latent Space Representation*

The effectiveness of the proposed supervised convolutional autoencoder (CAE) is illustrated through t-SNE projections of the EEG data before and after feature extraction. As shown in Fig. 6, raw features do not exhibit meaningful separation between rest and hand closure states. In contrast, the CAE learns compact and discriminative latent representations that clearly distinguish between the two motor classes. This indicates the CAE's ability to extract class-relevant features while suppressing irrelevant or noisy information, which is critical for improving classification performance in EEG-based motor decoding tasks

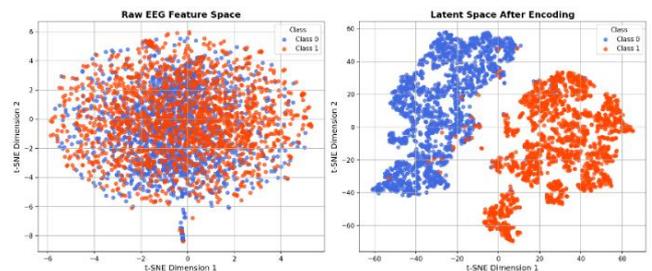

Figure.6. *T-SNE comparison of raw EEG (left) and CAE-extracted features (right), showing improved separation between rest (Class 0) and hand closure (Class 1) in the latent space.*

*C. Classification Model Performance*

To identify the model with the best generalization capability, we prioritized validation and test accuracy while considering the gap from training accuracy to avoid overfitting. Among the evaluated models, AdaBoost, ShallowFBCSPNet, and GRU demonstrated superior generalization performance. AdaBoost achieved the highest test accuracy (0.8930) with slightly lower training accuracy (0.9997), indicating better generalization than overfitted models like LightGBM. ShallowFBCSPNet and GRU also exhibited balanced performance, with high validation and test accuracies ( ≥ 0.8904) and moderate training accuracy, making them strong neural alternatives. Overall, AdaBoost showed the most favorable trade-off between performance and generalization, as shown as figure7.

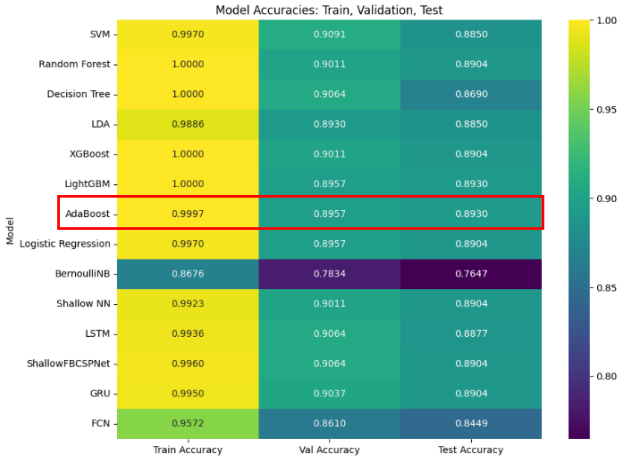

Figure.7. *Classification accuracy of various machine learning models on training, validation, and test sets*

## V. DISCUSSION

### A. Offline Evaluation

The results from offline analysis underscore the efficacy of the proposed supervised feature extraction pipeline. The convolutional autoencoder (CAE) demonstrated a strong ability to reduce data dimensionality while preserving discriminative power. Compared to raw features, the CAE-generated latent space allowed for clear separation between motor states, as evidenced by the t-SNE projections. This latent encoding not only improved classification performance but also enhanced robustness to noise and subject variability, critical aspects for EEG-based BCI systems. Among the classifiers evaluated, AdaBoost achieved the highest generalization performance accuracy with minimal overfitting (89.30%), outperforming deeper neural networks such as GRU and ShallowFBCSPNet. This highlights that, for the compact latent features learned through supervised encoding, simpler ensemble methods can offer effective decision boundaries while maintaining computational efficiency. The class-specific spatial distribution maps provided further insight into the neural basis of classification. The rest condition (Class 0) exhibited consistent spatial patterns across the central and fronto-central regions, particularly near sensorimotor areas, which are known to be involved in hand movement suppression and relaxation. These stable topographies confirm the neurophysiological grounding of the extracted features and support the validity of using EEG signatures for differentiating subtle motor states.

### B. Real-Time System Performance

Real- The proposed brain–computer interface (BCI) system was validated in a real-time setting as shown as figure 8 making flexion and extension .it was tested on five healthy participants (three males, two females), each performing 15 trials of motor imagery tasks involving hand opening and closing. As presented in Fig. 9, classification accuracies ranged from 60% to 86%, reflecting subject-specific variability in EEG signal quality and motor control performance. Notably, true positive rates consistently exceeded false positives, indicating a reliable detection of intentional motor imagery and confirming the robustness of the system in real-time operation. The same signal processing pipeline employed during offline analysis was retained for real-time implementation. This included data preprocessing, feature extraction, and classification, ensuring consistency across experimental stages. AdaBoost was selected as the final classifier due to its superior generalization performance in offline evaluations, achieving an accuracy of 89.30% with minimal overfitting.

To contextualize the system's performance, a comparative analysis with existing real-time BCI systems for hand movement decoding is provided in Table 1. The proposed system demonstrates competitive performance in terms of accuracy and responsiveness, validating its feasibility for practical neurorehabilitation applications.

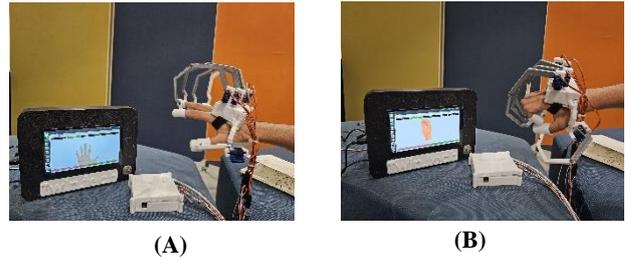

Figure.8. *BCI-driven hand exoskeleton. (A) Rest state with open hand; (B) Decoded command triggers hand closure.*

TABLE I. SUMMARY OF REAL-TIME BCI SYSTEMS FOR HAND MOVEMENT DECODING.

| Authors | No Subjects | Accuracy % | Headset / Channels |
|---|---|---|---|
| Arpa Suwannarat et al. [17] | 11 | 48.7 % | G.Nautilus / 16 |
| Susanjeewa Dharmasena et al. [18] | 8 | 70.38 % | Emotiv Epoc / 14 |
| Han Wei Ng et al. [19] | 50 | 61.05% | actiCHamp /32 |
| **Our Proposed Approach** | 5 | 89.30% (Offline), 60–86% (Real-time) | Emotiv Epoc /12 |

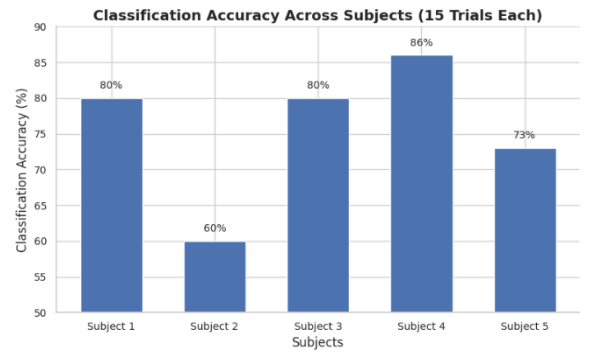

Figure.9. *Real-time classification accuracy across five subjects using a 12-channel Emotive EPOC EEG headset*

## C. Limitations and Future Directions

We only tested five healthy adults on a simple hand task with a short calibration. Results may change across devices because signal quality and protocols differ. We also didn't measure comfort/usability. Next, we'll run a pilot with stroke patients, broaden the gesture set (finger and wrist), and add a brief guided calibration with artefact- and confidence-gated decisions. We'll report end-to-end latency (mean and P95), safety, usability, and power/energy per decision, and compare against strong baselines using the same data splits.

## Conclusion

This study presents a fully integrated EEG-based BCI system for real-time neurorehabilitation, featuring a 3D-printed robotic exoskeleton that provides active motor assistance based on decoded brain signals. The system employs a supervised convolutional autoencoder for efficient and discriminative feature extraction from EEG recordings, enabling accurate classification of finger flexion versus rest. AdaBoost was selected for deployment due to its superior offline test accuracy (89.3%) and consistent real-time performance across five healthy participants. The system's seamless integration—including EEG acquisition, real-time classification, and robotic actuation—delivers reliable, low-latency neurofeedback via a compact and portable setup. These results validate the feasibility of the proposed BCI-exoskeleton solution for individualized rehabilitation. Future work will extend to more diverse stroke subjects, additional motor functions, and clinical testing to enhance the translational impact of the system.

## Acknowledgment

The authors gratefully acknowledge the Academy of Scientific Research and Technology (ASRT), Egypt, and Technological Incubator for Prostheses and Orthoses (TIPO) at the Military Technical College in Egypt for their generous support and funding of this research. Their commitment to translational innovation in assistive and rehabilitation technologies was instrumental to the success of this work.